\documentstyle[twocolumn,amssymb,aps,psfig,float]{revtex}
\begin{document}
\bibliographystyle{plain}
%\hskip -2\wd0\copy1
\twocolumn[\hsize\textwidth\columnwidth\hsize\csname @twocolumnfalse\endcsname

\title{ 
Magnetic properties of the coupled ladder system MgV$_2$O$_5$
}
\vskip0.5truecm 
\author{P. Millet$^1$, C. Satto$^1$, J. Bonvoisin$^1$, B. Normand$^2$, K.
Penc$^3$, M. Albrecht$^4$ and F. Mila$^4$}

\address{$^1$
	CEMES, CNRS, 29, rue J. Marvig, 31055 Toulouse Cedex, France. }
      
\address{$^2$Theoretische Physik, ETH-H\"onggerberg, CH-8093 Z\"urich, 
Switzerland. }

\address{$^3$Max Planck Institut f\"ur Physik komplexer Systeme, 
Bayreuther Str. 40, D-01187 Dresden, Germany. }

\address{$^4$Laboratoire de Physique 
Quantique, Universit\'e Paul Sabatier, 118 Route de Narbonne, 31062 
Toulouse Cedex, France. }\vskip0.5truecm
      
\maketitle

\begin{abstract}
\begin{center} 
\parbox{14cm}{
We present magnetic susceptibility measurements on MgV$_2$O$_5$, a compound 
in which the vanadium oxide planes have the same topology as in CaV$_2$O$_5$.
The most striking property is that there is an energy gap of about 15 K, much
smaller than in CaV$_2$O$_5$, where the values reported are of the order of
500 K. We show that this may be understood naturally in terms of the phase
diagram of the frustrated, coupled ladder system. This analysis leads to the
prediction that 
MgV$_2$O$_5$ should have incommensurate dynamic spin fluctuations. 
}
\end{center}
\end{abstract}
\vskip .1truein
 
\noindent PACS Nos : 75.10.jm 75.40.Cx 75.50.Ee
\vskip2pc]

The study of low dimensional, spin-1/2 magnets is a very active field of
research. One of the recent successes has been the synthesis and theoretical
understanding of spin ladder systems\cite{dagotto}. These consist of 
weakly interacting pairs of antiferromagnetic chains coupled by 
antiferromagnetic rungs, and the single most interesting result is the 
appearance of a gap in the spin excitation spectrum. The
prototype of such systems is SrCu$_2$O$_3$, whose magnetic properties can be
accounted for by a spin-1/2 Heisenberg model
\begin{equation}
H = \sum_{(i,j)} J_{ij} \vec S_i . \vec S_j
\end{equation}
restricted to the exchange integrals depicted in Fig. 1. In the case of 
SrCu$_2$O$_3$, it is believed that $|J_1| \ll J_2, J'_2$, so that the system 
is clearly in the ladder limit. The family of planar vanadates also contains 
compounds where V atoms in ionization state V$^{4+}$ give rise to spin-1/2 
planes having the topology of Fig. 1, and their primary difference from 
SrCu$_2$O$_3$ is that the coupling between the ladders
is not {\it a priori} a small parameter. The first example in which a spin 
gap was reported is CaV$_2$O$_5$\cite{iwase}. The presence of a 
large spin gap (about 500 K) in spite of non-negligible inter-ladder 
coupling came as a surprise, and prompted four of
the present authors to study the phase diagram of the model depicted in Fig.
1\cite{normand}. In particular, they showed that the ladders behave in an 
essentially decoupled manner when the second-neighbor coupling constants 
$J_2$ and $J_2'$ are large enough compared to $J_1$, and that there 
should be an ordered, incommensurate phase when the three coupling constants 
are comparable. These results provide a reasonable
explanation of the presence of a large spin gap in CaV$_2$O$_5$ since, as we
will argue below, the information currently available on the magnetic 
interaction parameters leads to the estimates $J_2/J_1 \simeq 3.5$ and 
$J_2'/J_1 \simeq 4.3$ for that system. Materials which could be used to 
probe other regions of the rather rich phase diagram of Ref. 
\cite{normand} would be very welcome.

In this paper, we present magnetic susceptibility measurements for 
another member of the vanadate family with the same topology, namely 
MgV$_2$O$_5$. This system is quite intriguing because, although the 
vanadium oxide planes are very similar to those in CaV$_2$O$_5$\cite{bouloux1}, 
the magnetic
properties are strikingly different, with, in particular, a spin gap of only 
15 K. As we shall argue below, this difference may be attributed to
the presence of a helical ordered phase in the phase diagram of the model of
Fig. 1, thus lending further support to the results of Ref. \cite{normand}. 

MgV$_2$O$_5$ was prepared by solid state reaction, starting from a mixture of 
MgO
and VO$_2$, according to the procedure described in Ref. \cite{bouloux}, and its
structure was determined from x-ray powder diffraction Rietveld
analysis\cite{millet}. It crystallizes in the orthorhombic system of space group
{\it Cmcm} with, at 294 K, $a$ = 3.6913(2)\AA , $b$ = 9.9710(4)\AA , and $c$ =
11.0173(4)\AA . The structure retains the basic framework of the V$_2$O$_5$ host
lattice, which is built up in one direction by infinite double lines of VO$_5$
square pyramids joined by alternate sharing of edges and corners to form 
vanadium
zigzag chains along the $a$ axis, and held together by corner sharing in 
the [001]
direction. 
It is therefore indeed analogous to the structure of the
compounds $\alpha'$-NaV$_2$O$_5$\cite{hardy} and CaV$_2$O$_5$\cite{bouloux1}. 
A peculiar feature of the
MgV$_2$O$_5$ structure is the shift by $a/2$ along [100] for alternate layers
(doubling of the short inter-layer parameter), and the puckering of the
V$_2$O$_5$ layers (angle $\mu = 21^o$) to accommodate Mg ions in tetrahedral
coordination. An idealized view of the structure viewed down the $a$ axis is 
presented in Fig. 2. As for CaV$_2$O$_5$, MgV$_2$O$_5$ is a monovalent 
(V$^{4+}$) oxide, and vanadium atoms are located on a single crystallographic 
site.

Magnetic susceptibility measurements were performed using a superconducting
quantum interference device QUANTUM DESIGN SQUID susceptometer. The magnetic
field induction was 1kG. The molar susceptibility was corrected for diamagnetism
by using Pascal's constants. The raw data are presented in Fig. 3. These data
were reproduced on three different samples, with, however, a large variation 
of the temperature-independent background from sample to sample. 
The main features of these data are: i) a regular increase between 400 K and 
100 K; ii) a small, not fully reproducible bump around 85 K; iii) a very slow 
decrease between 70 K and 15 K; iv) a much faster decrease down to 4 K; v) a
very fast increase at lower temperature. 

The low-temperature data are typical of a system with a spin gap. We have thus
performed a fit of the data between 2 K and 25 K with the formula
\begin{equation}
\chi(T)=A + B/T + (C/\sqrt T) \exp(-\Delta/T),
\end{equation}
where $A$ includes all temperature-independent contributions (Van Vleck, plus
possible contributions from impurity phases with temperature-independent
susceptibility in this range), $B/T$ accounts for paramagnetic impurities, and 
$(C/\sqrt T) \exp(-\Delta/T)$ describes the susceptibility of a 
one-dimensional system with a gap\cite{troyer}. The fit is depicted in the 
inset of Fig. 3, and the parameters used are $A = 1.71\, 10^{-3}\, 
{\rm emu/mole}$, 
$B = 1.16\, 10^{-3}\, {\rm emu K/mole}$, 
$C = 7.28\, 10^{-3}\, {\rm emu K^{1/2}/mole}$ and $\Delta = 14.8 K$. The fit is 
very
accurate, with a residue of only $ 1.6 \times 10^{-6}$ over a temperature range
2 - 25 K. We have also tried 
to fit the data with a term of the form $C \exp(-\Delta/T)$ typical of gapped, 
two-dimensional systems\cite{troyer}. The result was clearly not as good, which 
we take to indicate the effective one-dimensionality of the system, or that 
MgV$_2$O$_5$ remains in the limit of weakly interacting ladders.

While further investigation of this system, for example by Nuclear Magnetic 
Resonance, would be very useful to confirm these results, the presence of a 
rather small gap ($\simeq 15 K$) as compared to CaV$_2$O$_5$ ($\simeq 500 K$) 
is clearly very interesting. We believe that this result can be understood 
in terms of the model of Fig. 1, due to modest but not negligible differences 
in the structural parameters leading to important differences in the
exchange integrals. The direct determination of exchange integrals for a
particular system is a very difficult task, because it requires an accurate fit
of the temperature-dependence of the susceptibility over a large temperature
range. In vanadates this has been possible only in two cases for which we 
have a good theoretical understanding of the temperature-dependence of the 
susceptibility of the underlying model, 
$\alpha'$-NaV$_2$O$_5$\cite{mila,isobe2} 
and CsV$_2$O$_5$\cite{isobe}. $\alpha'$-NaV$_2$O$_5$ is a one-dimensional 
Heisenberg chain of corner-sharing VO$_5$ pyramids with a single exchange 
integral $J_2$(NaV$_2$O$_5$). The susceptibility of this model is known very 
accurately from the work of many authors\cite{eggert}, and a fit
of the experimental susceptibility gives $J_2$(NaV$_2$O$_5$)$\simeq 530 K$.
CsV$_2$O$_5$ consists of weakly interacting pairs of edge-sharing pyramids. The
relevant model is a Heisenberg model with only two spins coupled by an exchange
integral $J_1$(CsV$_2$O$_5$). The
susceptibility can be calculated analytically, and a fit to the experimental
results gave $J_1$(CsV$_2$O$_5$)$\simeq 146 K$\cite{isobe}. 

To obtain estimates of the exchange integrals in other members of the family,
we make the reasonable assumption that these depend only on the local 
geometry of
the bonds. For corner-sharing pyramids ($J_2$,$J_2'$), the dominant process 
is superexchange, 
and the exchange integrals are expected to depend on the local parameters 
defined in Fig. 4 according to\cite{mila} 
\begin{equation}
J_2(d, \theta), J_2'(d, \theta) \propto \cos^4(\theta) d^{-14},
\end{equation}
where the exponent is given by empirical laws concerning the dependence of 
the overlap integrals on distance\cite{harrison}.
For edge--sharing pyramids ($J_1$), the origin of the interaction is not 
so clear. Assuming that 
it is due primarily to direct exchange between vanadium 3d orbitals, similar 
empirical laws predict that it should scale according to
\begin{equation}
J_1(d_{V-V}) \propto  (d_{V-V})^{-10}.
\end{equation}
where $d_{V-V}$ is the vanadium-vanadium distance.
Having no other, more direct source of information at hand for vanadates, 
we will use these dependences in the following. The results for
the exchange integrals of CaV$_2$O$_5$ and MgV$_2$O$_5$ based on the most 
accurate structural information
available for $\alpha'$-NaV$_2$O$_5$\cite{carpy}, CsV$_2$O$_5$\cite{isobe}, 
CaV$_2$O$_5$\cite{onoda} and
MgV$_2$O$_5$\cite{millet} are given in Table I. 
The trend is quite clear: $J_1$ is larger for MgV$_2$O$_5$
than for CaV$_2$O$_5$, while $J_2'$ is considerably smaller, so that the 
ratio $J_2'/J_1$ is approximately 4.3 for CaV$_2$O$_5$, but only 2.5 for 
MgV$_2$O$_5$. According to the analysis of the model performed in Ref.
\cite{normand}, this ratio is a very important parameter: $J_2'/J_1$ = 4.3 
clearly puts CaV$_2$O$_5$ in the ladder limit with a large gap\cite{note}. 
However, $J_2'/J_1$ = 2.5 with $J_2/J_1$ = 2.8 puts MgV$_2$O$_5$ 
very close to the helical
ordered - hence gapless - phase according to the phase diagram obtained by 
Schwinger boson mean-field theory\cite{normand}. 
The 
system is thus expected to have a very small gap, 
as this vanishes continuously at
the boundary. Of course, estimates based on Eqs. (3) and (4) are not very
accurate\cite{eskes}, but the conclusions can be expected to remain 
qualitatively correct as long as the exchange integrals depend strongly on 
the distances, which has indeed been found to be
the case in insulating oxides\cite{aronson}.

An interesting consequence of the proximity to a helical ordered phase is that 
MgV$_2$O$_5$ should exhibit incommensurate dynamic spin fluctuations. We do not 
present calculations to elaborate on this point here. This prediction
could be verified by inelastic neutron scattering when single crystals of
sufficient sizes become available.

Let us now turn to the high-temperature data. The increase of susceptibility 
with decreasing temperature, showing a maximum at intermediate temperatures, 
is typical of systems with antiferromagnetic exchanges. Fitting the measured 
curve with a Heisenberg model has proven to be as difficult here as for other 
members of the family. The problem of the peak occurring well below the energy 
scale of the interactions may be due to poor theoretical estimates of the 
temperature dependence for frustrated systems with dynamic helical 
fluctuations. Work is in progress in this direction. 

We focus finally on the broad susceptibility peak which occurs around 85 K. 
To address the possibility that this behavior is related to structural 
modifications, we have studied 
the thermal evolution of the x-ray diffractograms. X-ray powder patterns
were collected up to $\theta=30^o$ (except the diffractogram at 83 K which was
collected up to 75$^o$ in order to perform a Rietveld refinement\cite{millet})
in a high-accuracy MICROCONTROLE diffractometer, using CuK$\alpha$ radiation
(graphite monochromator) from an 18kW rotating-anode generator and a flowing
cryostat with a stability range of 0.5 K. The thermal variation of the cell
parameters, determined by least-squares refinement, is presented in Fig. 5. The 
lattice parameter $a$ along the vanadium chains increases slightly from 170 K 
to 100 K, a temperature at which the fluctuations in the data suggest that
a structural instability may be occurring, then decreases monotonically from
65 K to 8 K. Interestingly, the instability temperature range 100-65 K
corresponds to the maximum observed in the susceptibility curve. 
Moreover, a shortening of the lattice parameter $a$ is qualitatively
consistent with an 
increase
of the exchange integrals $J_1$ and $J_2$, hence with the experimental decrease
of the susceptibility. However, it is
difficult to correlate this with a structural modification because we did not
observe a notable change on the Rietveld refinement of the structure performed
at 83 K\cite{millet}. The interlayer distance $b$ decreases continuously 
over the
whole temperature range, while $c$ presents no clear evolution. A structural 
determination at low temperatures (around 4 K) would be very useful, and we 
plan to perform this in the near future by neutron powder diffraction.

To summarize, the magnetic susceptibility data reported in this paper suggest
that MgV$_2$O$_5$ is a new spin gap system with a gap of about 15 K. 
This very small value should be contrasted with the much larger value 
($\simeq $500 K) reported for CaV$_2$O$_5$. We have shown that this difference 
can be understood in terms of the frustrated, coupled ladder system due to the
proximity to an ordered helical phase in the case of MgV$_2$O$_5$. This 
interpretation suggests that MgV$_2$O$_5$ should have
incommensurate dynamic spin fluctuations, a prediction 
which could be tested by inelastic neutron scattering as soon as suitably 
large crystals are available. Given the potential interest of systems
with spin gaps in the range 0-30 K for studies under magnetic field, 
the magnetic properties of this system are likely to become a very active field 
of research.  

We would like to thank P. Sciau for his help with the x-ray diffraction
analysis, and also acknowledge useful discussions with J.-P. Daudey and 
J. Galy.

\begin{table} [h]
\begin{center} 
\begin{tabular}{lccccccccc}
  & $d_{V-V}$ & $J_1$ & $d_\perp$ & $d$($J_2$) & $d$($J_2'$) & $J_2$  & $J_2'$  
  & $J_2/J_1$ & $J_2'/J_1$ \\
\hline
CsV$_2$O$_5$ & 3.073 & 146 &  & & & & & & \\
\hline
$\alpha'$-NaV$_2$O$_5$ & & & 0.698 & 1.962&  & 530 & & & \\
\hline
CaV$_2$O$_5$ & 3.0257 & 170 & 0.648 & 1.949 & 1.905 & 587 & 730 & 3.5 & 4.3 \\
\hline
MgV$_2$O$_5$ & 2.976 & 201 & 0.666 & 1.954 & 1.975 & 565 & 511 & 2.8 & 2.5 \\
\end{tabular}
\vskip.5cm
\caption{
Structural parameters and measured or estimated exchange integrals of
CsV$_2$O$_5$, $\alpha'$-NaV$_2$O$_5$, CaV$_2$O$_5$ and MgV$_2$O$_5$.
Distances are in \AA, and  exchange integrals are in Kelvin.}
\end{center}
%\label{tableau}
\end{table}
 
\begin{figure}
\centerline{\psfig{figure=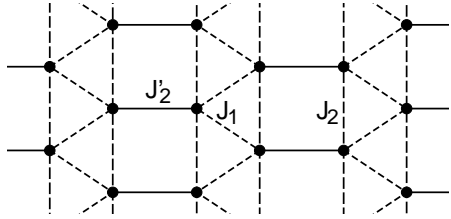,width=6.0cm,angle=0}}
\caption{Schematic representation of the frustrated coupled ladder system with
one type of nearest neighbour exchange integral ($J_1$) and two types of next
nearest neighbour exchange integrals ($J_2$,$J_2'$)
}
\end{figure}

\begin{figure}[hp]
\caption{Idealized structure of MgV$_2$O$_5$ viewed down the a axis (3.69\AA ).
}
\end{figure}

\begin{figure}
\centerline{\psfig{figure=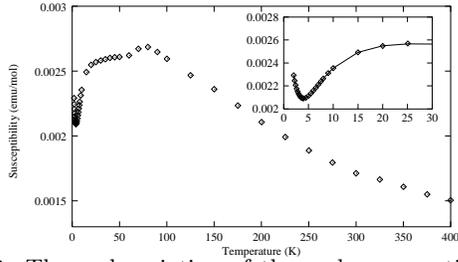,width=6.0cm,angle=-90}}
\caption{Thermal variation of the molar magnetic susceptibility of 
MgV$_2$O$_5$. Diamonds: experimental points. Inset: enlargement of the 
low-temperature region with fit (solid line) to Eq. (2).
}

\end{figure}

\begin{figure}
\caption{ a) V-$3d$ and O-$2p$ orbitals involved in superexchange between 
corner-sharing VO$_5$ square pyramids; b) basic parameters of the VO$_5$ 
square pyramid.
}
\end{figure}

\begin{figure}
\caption{Temperature dependence of lattice constants for MgV$_2$O$_5$. Triangles
and black dots correspond to two successive collection sets. The open circles
correspond to the Rietveld refinement at 83 K.
}
\end{figure}


\begin{references}

\bibitem{dagotto} See e.g. E. Dagotto and T. M. Rice, Science {\bf 271}, 618
(1996), and references therein.
\bibitem{iwase} H, Iwase, M. Isobe, Y. Ueda and H. Yasuoka, J. Phys. Soc. Jpn.
{\bf 65}, 2397 (1996).

\bibitem{normand} B. Normand, K. Penc, M. Albrecht and F. Mila, Phys. 
Rev. B {\bf 56}, R5736 (1997).

\bibitem{bouloux1} See J.-C. Bouloux and J. Galy, J. Solid State Chem. {\bf 16},
385 (1976), and references therein.

\bibitem{bouloux} J.-C. Bouloux, I. Milosevic and J. Galy, J. Solid State Chem.
{\bf 16}, 393 (1976).

\bibitem{millet} P. Millet, C. Satto, P. Sciau and J. Galy, unpublished.

\bibitem{hardy} 
A. Hardy, J. Galy, A. Casalot and M. Pouchard, Bull. Soc. Chim. Fr.
{\bf 4}, 1056 (1965);
J. Galy, A. Casalot, M. Pouchard, P. Hagenmuller, C. R. Acad. Sc.
{\bf 262 C}, 1055 (1966);
M. Pouchard, A. Casalot, J. Galy et P. Hagenmuller, Bull. Soc.
Chim. Fr. {\bf 11}, 4343 (1967).

\bibitem{troyer} M. Troyer, H. Tsunetsugu and D. W\"urtz, Phys. Rev. B {\bf 50},
13515 (1994).

\bibitem{mila} F. Mila, P. Millet and J. Bonvoisin, Phys. Rev. B {\bf 54}, 
11925
(1996). 

\bibitem{isobe2} M. Isobe and Y. Ueda, J. Phys. Soc. Jpn. {\bf 65}, 1178 (1996).

\bibitem{isobe} M. Isobe and Y. Ueda, J. Phys. Soc. Jpn. {\bf 65}, 3142 (1996).

\bibitem{eggert} See S. Eggert, I. Affleck and M. Takahashi, Phys. Rev. Lett. 
{\bf
73}, 332 (1994), and references therein.

\bibitem{harrison} W. A. Harrison, {\it Electronic Structure and the 
Properties of
Solids} (Freeman, San Francisco, 1980).

\bibitem{carpy} A. Carpy and J. Galy, Acta Cryst. Sect. B {\bf 31}, 1481
(1975).

\bibitem{onoda} M. Onoda and N. Nishiguchi, J. Solid State Chem. {\bf 127},
359 (1996).

\bibitem{note} Precise estimates of the gap in that limit are difficult to
obtain, but it is expected to be around 0.6 $J_2'$, in reasonable
agreement with the experimental estimate.

%\bibitem{note2} According to Schwinger boson mean-field theory, the N\'eel 
order
%for $J_2=0$ is stable until $J_2'/J_1=2.65$, while the essentially exact result
%of Monte Carlo simulations is $J_2'/J_1=1.74$.

\bibitem{eskes} H. Eskes and J. Jefferson, Phys. Rev. B {\bf 48}, 9788 (1993).

\bibitem{aronson} M. C. Aronson et al,  Phys. Rev. B {\bf 44}, 4657 (1991).

\end{references}
\end{document}